# First-order disorder-driven transition and inverse melting of the vortex lattice

Nurit Avraham[a], Boris Khaykovich[a], Yuri Myasoedov[a],
Michael Rappaport[a], Hadas Shtrikman[a], Dima E. Feldman[a,1], Eli Zeldov[a],
Tsuyoshi Tamegai[b], Peter H. Kes[c], Ming Li[c], Marcin Konczykowski[d], and Kees van der Beek[d]

*a Department of Condensed Matter Physics, The Weizmann Institute of Science, Rehovot 76100, Israel*
*1 Materials Science Division, Argonne National Laboratory, Argonne, IL 60439, and also at Landau Institute for Theoretical Physics, 142432 Chernogolovka, Moscow region, Russia*
*b Department of Applied Physics, The University of Tokyo, Hongo, Bunkyo-ku, Tokyo 113-8656, and CREST, Japan Science and Technology Corporation (JST), Japan*
*c Kamerlingh Onnes Laboratory, Leiden University, 2300 RA Leiden, The Netherlands*
*d Laboratoire des Solides Irradies, CNRS UMR 7642 and CEA/DSM/DRECAM, Ecole Polytechnique, 91128 Palaiseau, France*

**Abstract**

Vortex matter phase transitions in the high-temperature superconductor $Bi_2Sr_2CaCu_2O_8$ were studied using local magnetization measurements combined with a vortex 'shaking' technique. The measurements revealed thermodynamic evidence of a first-order transition along the second magnetization peak line, at temperatures below the apparent critical point $T_{cp}$. We found that the first-order transition line does not terminate at $T_{cp}$, but continues down to at least 30 K. This observation suggests that the ordered vortex lattice phase is destroyed through a unified first-order transition that changes its character from thermally induced melting at high temperatures to a disorder-induced transition at low temperatures. At intermediate temperatures the transition line shows an upturn, which implies that the vortex matter displays 'inverse' melting behavior.

## 1. Introduction

Vortex matter in high-temperature superconductors (HTS) has a rich and complicated phase diagram. The nature of the different vortex matter phases and the transitions between them are the subjects of intense theoretical and experimental investigations. It is generally accepted that at high temperatures the relatively ordered vortex lattice melts into a vortex liquid phase via a *thermally induced* first-order transition (FOT) [1]-[5] whereas at low temperatures the ordered vortex lattice transforms into a disordered vortex phase via a *disorder-driven* transition characterized by the second magnetization peak [6]-[8]. It was proposed that the first-order melting line terminates at a critical point $T_{cp}$ that defines the crossover from the FOT to the second peak regime.

While the high temperature region of the phase diagram has been extensively studied, at low and intermediate temperatures the investigation of the equilibrium vortex state is difficult due to pinning of vortices by material disorder. Hence, one of the major open questions in the phase diagram of HTS is the thermodynamic nature of the disorder-driven transition. Another important issue that is not well understood is the flattening of the melting line near the crossover to the disorder-driven transition and the termination of the thermal melting line at the critical point.

In this work we have investigated the phase diagram of $Bi_2Sr_2CaCu_2O_8$ (BSCCO) crystals at low temperatures. We have focused on the critical point and on the behavior of the melting line in the vicinity of $T_{cp}$, as well as on the thermodynamic nature of the disorder-driven transition. In particular we addressed the question whether this transition is of a first-order or continuous nature. In order to overcome the difficulties arising from the non-equilibrium conditions of low temperatures, we implemented a recently introduced 'shaking' technique, using an in-plane *ac* magnetic field [9],[10]. This technique was found to depin the vortex lattice, enabling the observation of equilibrium vortex properties even at low temperatures.

## 2. Experimental

The experiments were carried out on a number of optimally-doped BSCCO crystals ($T_c \approx 90$ K) grown in two laboratories [11],[12]. We present here results on three crystals: A-160×600×25 µm$^3$, B-190×900×30 µm$^3$, and C-170×600×20 µm$^3$. The samples were mounted onto the surface of an array of eleven Hall sensors of 10x10 µm$^2$ and 10 µm apart (see Fig. 1). The active layer of the sensors is a two-dimensional electron gas (2DEG) formed at a GaAs/AlGaAs interface. The 2DEG is only ≈ 1000 Å below the surface, resulting in a very accurate measurement of the local magnetic field at each of the sensor locations across the sample. The 2DEG has a mobility of about $1 \times 10^5$ cm$^2$/V sec at 80 K and a density of about $6 \times 10^{-11}$ cm$^{-2}$ resulting in a sensitivity of about 0.1 Ω/Gauss. A 50 µA current was driven through the sensor array and the Hall voltage was measured using a Keithley 182 digital voltmeter for opposite polarities of the measuring current. All the sensors in the array were measured sequentially at each temperature and field by switching the voltmeter connection to each sensor, using a Keithley 7001 switch system.

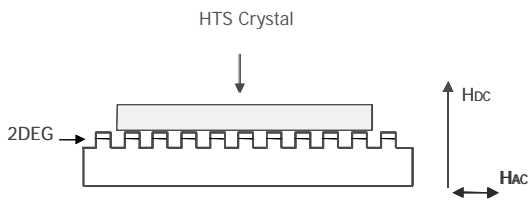

Fig. 1 Schematic side view of BSCCO crystal glued on the surface of the Hall sensors

The transverse *ac* field $H_{ac\perp}$ was generated by a small Helmholtz coil mounted on the sample rod, within the core of a superconducting coil. The Helmholtz coil was wound from Cu wires, with a current response of 80 G/A. The *ac* field was parallel to both the surface of the sample and the sensor array. The magnetic field $H_a$ parallel to the c-axis and the in-plane *ac* field were applied simultaneously to the sample. An audio amplifier driven by the sine output of a lock-in amplifier, provided the *ac* current to the coil.

The efficiency of the vortex shaking depends on the frequency and amplitude of the *ac* magnetic field. Experimentally, a frequency of 1 kHz was found to be optimal. The amplitude of $H_{ac\perp}$ was varied according to the measurement temperature – an *ac* RMS current of 450 mA (36 G RMS) was sufficient to eliminate the irreversibility of the magnetization loop at temperatures above 50 K. At intermediate temperatures, between 38 K and 50 K, we usually applied an *ac* current of about 700 mA. At low temperatures we used 1-1.3 A; above 1.3 A the heat produced by the coil prevented satisfactory temperature stabilization.

## 3. Eliminating irreversibility using the in-plane *ac* field

One of the major obstacles encountered when trying to study the behavior of the vortex matter at low temperatures is the formation of non-equilibrium states due to vortex pinning that conceal the thermodynamic phase transitions. In such non-equilibrium conditions, the sample shows an irreversible magnetization curve. In YBa$_2$Cu$_3$O$_7$, at temperatures close to $T_c$, 'vortex shaking' by a transverse *ac* field was shown to reduce the irreversible magnetization caused by vortex pinning [9]. We find that in BSCCO crystals this method can fully suppress the magnetization hysteresis even at low temperatures down to 30 K [10], as shown in Fig. 2a. The figure shows two local magnetization loops at $T = 30$ K, one with the additional *ac* field and the other without it. The measurement without the *ac* field shows hysteresis and the characteristic second magnetization peak, whereas the measurement with $H_{ac\perp}$ displays *fully reversible* magnetization.

The Abrikosov vortices in BSCCO can be regarded as a stack of Josephson-coupled pancake vortices [13],[14] in the individual $CuO_2$ planes. The in-plane field $H_{ac\perp}$ readily penetrates through the sample [2] in the form of Josephson vortices (JV) residing in-between the $CuO_2$ planes [15]. The main effect, which is of interest here, is that a pancake vortex, located in a $CuO_2$ plane immediately above a JV, is displaced a small distance along the direction of the JV relative to the neighboring pancake vortex residing one $CuO_2$ layer underneath [15]. For small values of $H_{ac\perp}$ the probability of such a close proximity between a pancake and a JV is low. Thus, in our sample geometry, an average pancake experiences only a few such periodic 'intersections' during one *ac* cycle with a duration of each intersection of less than 1% of the *ac* period. $H_{ac\perp}$ therefore induces a weak local *ac* agitation of pancakes, which assists thermal activation in reducing the irreversible magnetization and in approaching thermal equilibrium. In addition to bulk pinning, a substantial part of the magnetic hysteresis in BSCCO crystals is caused by surface [16] and geometrical [17] barriers. We found that $H_{ac\perp}$ also efficiently suppresses this source of hysteresis, probably through instantaneous cutting of the vortices by the Josephson vortices, allowing easy penetration of individual pancakes through the sample edges. The interaction between JV and pancake vortices in BSCCO and the suppression of irreversibility were recently also observed by direct imaging of the vortices [18].

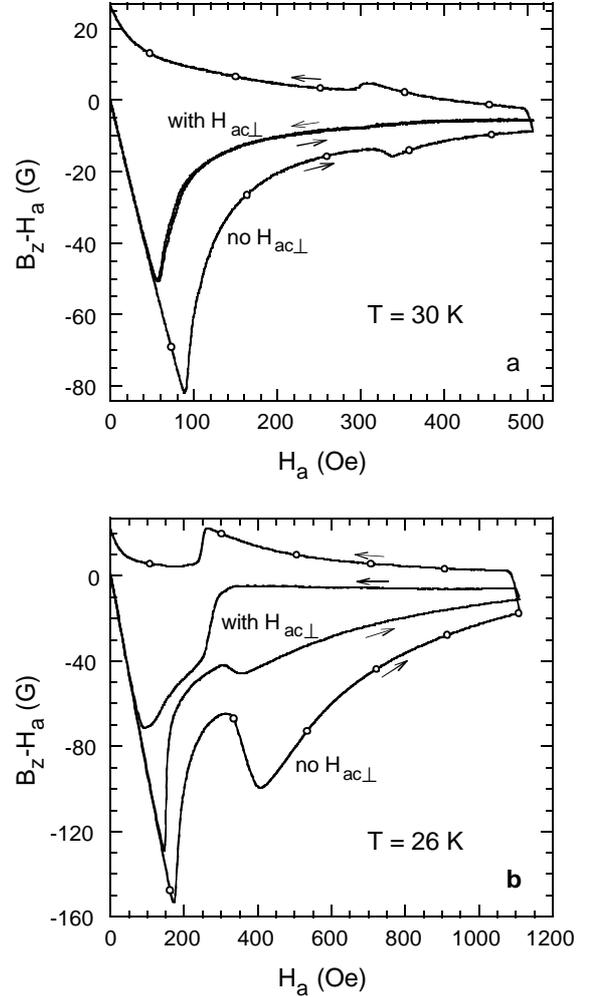

Fig. 2 Local magnetization loops in a BSCCO crystal with and without (O) transverse *ac* field $H_{ac\perp}$. (a) Crystal A at 30 K, the lowest temperature at which a fully reversible magnetization is attained for our maximum $H_{ac\perp}$ of about 80 G. (b) Crystal C at 26 K. Our maximum $H_{ac\perp}$ was not sufficient to obtain reversible magnetization.

Figure 2b shows an example of the magnetization loop at a lower temperature of 26 K. Here, the width of the *dc* loop is significantly larger than the width at 30 K and our maximum amplitude of $H_{ac\perp}$ was not sufficient to obtain reversible magnetization. It is interesting to note that the efficiency of $H_{ac\perp}$ is different in the two vortex phases. For the same width of the original loop, a much stronger suppression of the hysteresis of the ordered lattice was observed as compared to the disordered phase above the second magnetization peak. This is probably because the disordered phase exhibits stronger vortex pinning.

## 4. The thermodynamic nature of the phase transitions

A direct way to determine the nature of a phase transition is to observe the discontinuity in the density of vortices [1],[2]. At high temperatures, in clean BSCCO crystals, the first-order melting transition manifests itself as a clear and sharp step in the local magnetization curve [1],[19]. However, at lower temperatures this magnetization step was suppressed by the appearance of bulk pinning and irreversibility, and was found to disappear abruptly at a specific temperature that was regarded as a critical point. Eliminating the irreversibility by applying $H_{ac\perp}$ allowed us to observe the magnetization step at temperatures below this critical point.

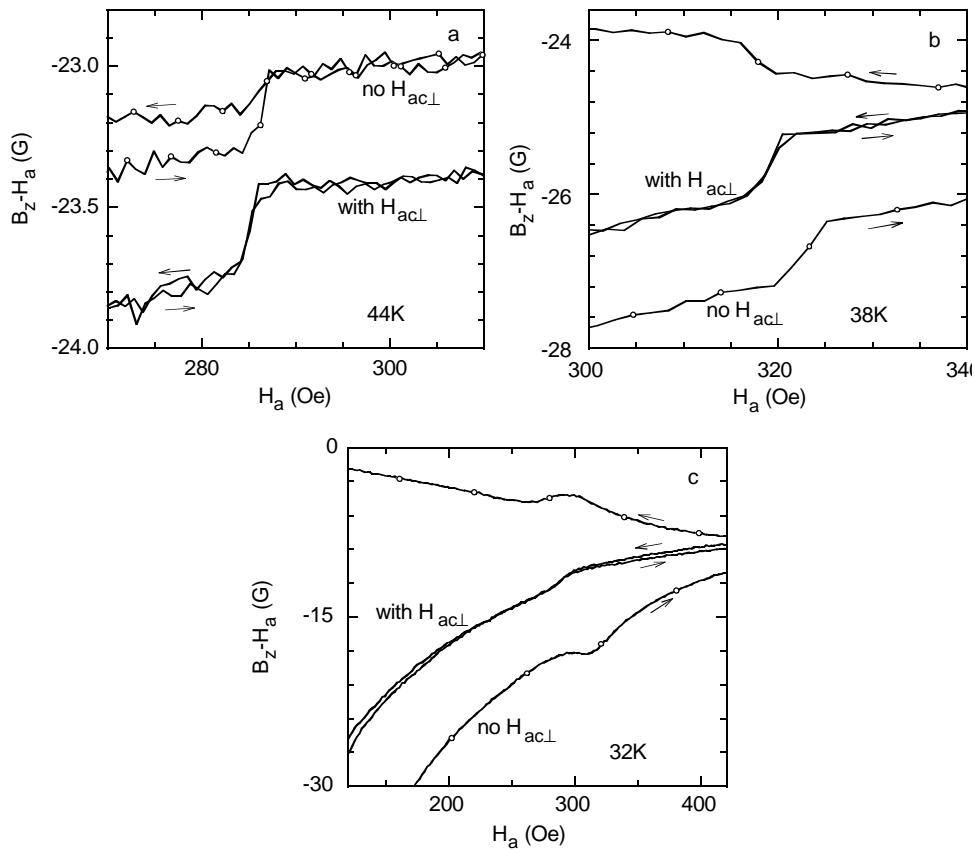

Fig. 3 Local magnetization loops in the vicinity of the melting transition with and without (O) $H_{ac\perp}$. (a) Crystal B at 44K, above the critical point. (b) Crystal B at 38K, just below the critical point. (c) Crystal C at 32K, far below the critical point.

Figure 3 shows local magnetization curves in the vicinity of the first-order melting transition at different temperatures. At 44 K (Fig. 3a), which is above $T_{cp} = 40$ K in this crystal, the magnetization step can still be resolved in the curve measured without the $H_{ac\perp}$, but is partially obscured by hysteresis that develops in the vortex solid just below the transition. Application of $H_{ac\perp}$ fully removes the hysteresis and enhances the magnetization step, as seen in the curve measured with $H_{ac\perp}$. Figure 3b demonstrates the behavior at 38 K, just below $T_{cp}$. At this temperature, the second magnetization peak and large hysteresis start to develop. The application of $H_{ac\perp}$ removes the hysteresis and reveals the underlying magnetization step, indicating the first-order nature of the transition.

A similar effect is obtained at 32K – far below the critical point – as shown in Fig. 3c. Here a fully developed second magnetization peak turns into a step in the reversible magnetization loop upon application of $H_{ac\perp}$. Note that the magnetization step is much smoother in this figure; this is a different crystal which displayed a broader step in the whole temperature range. Some broadening of the magnetization step is apparent in all figures. We attribute this broadening to the transverse field, which is known to slightly shift the melting field [20]. This periodic shift during the *ac* cycle should cause smearing of the magnetization step.

In contrast to previous observations – in which the magnetization step disappears abruptly at $T_{cp}$, replaced by a second magnetization peak below $T_{cp}$ – our results show that when applying $H_{ac\perp}$ the magnetization step continues down to much lower temperatures. This suggests that the FOT does not terminate at $T_{cp}$, but rather extends far below. An interesting question is whether the mechanism of the transition at $T < T_{cp}$ is the same as at higher temperatures. One of the ways to address this issue is to determine the shape of the FOT curve on the *B-T* phase diagram near $T_{cp}$. If thermal fluctuations are the main source of the FOT, the transition line should extend upwards with decreasing temperature. On the other hand, if the transition is disorder driven it should flatten out and follow the second magnetization peak line.

In order to derive the *B-T* phase diagram, we need to determine accurately the field at which the transition occurs. This was done by analyzing the derivative of the magnetization with respect to the induction $B_z$ in the vicinity of the transition. Figure 4 shows the derivative of $B_z$ - $H_a$ with respect to $B_z$, as a function of $B_z$, in the temperature range between 32 K and 44 K, in the presence of the $H_{ac\perp}$. The transition field $B_m$ was determined as the peak of the Gaussian-like derivative curve, which is the value of the field at which the magnetization has the steepest slope. Using this procedure, we were able to map the transition line very accurately on the *B-T* phase diagram. Other criteria result in a very similar curve.

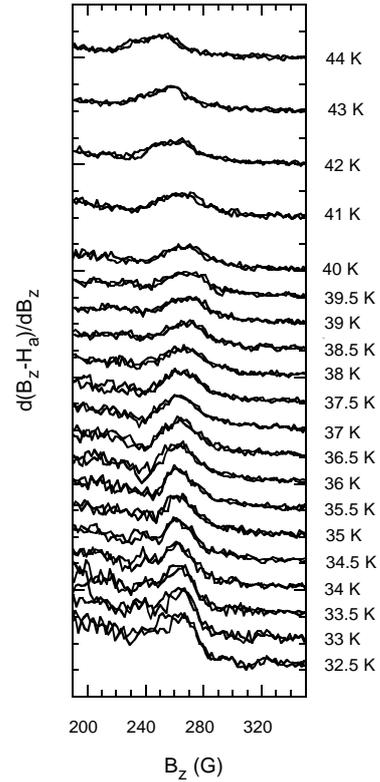

Fig. 4 Derivative of the magnetization $B_z$-$H_a$ with respect to $B_z$, as a function of $B_z$ in crystal B. Every curve corresponds to a different temperature in the range from 32.5 K to 44 K, as indicated on the right axis. The curves are vertically displaced for clarity.

Figure 5a shows a magnified view of the melting line on the *B-T* phase diagram in the vicinity of the proposed $T_{cp}$. The termination of the FOT line at 31 K indicates the lowest temperature below which we could not achieve reversible magnetization loops.

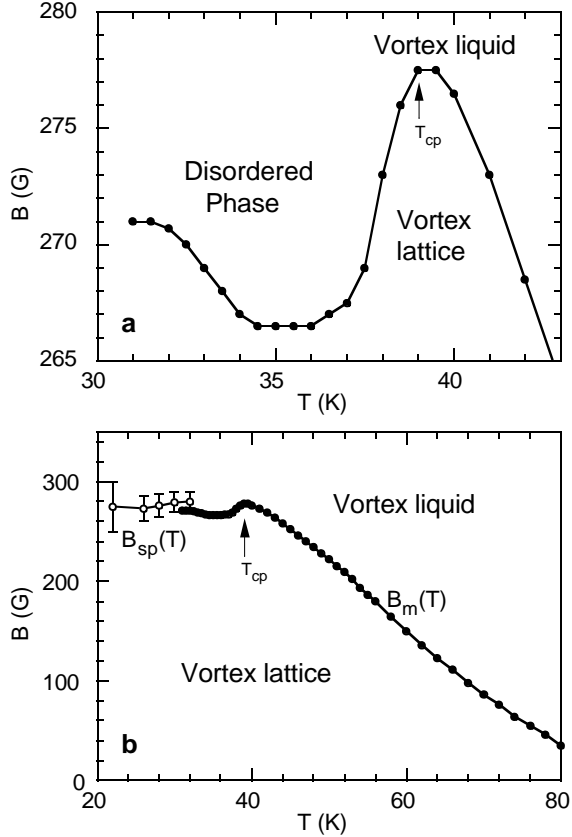

Fig. 5 (a) Magnified view of the melting line on a *B-T* phase diagram in the vicinity of the critical point in crystal B. At high temperatures, the vortex lattice transforms into a vortex liquid phase through a thermally induced FOT whereas at lower temperatures the vortex lattice transforms into a disordered vortex phase through a disorder-driven FOT. Below $T_{cp}$ the line shows a positive slope $dB_m/dT$, displaying inverse melting behavior. (b) The entire melting transition $B_m(T)$ along with the second peak transition $B_{sp}(T)$ (O).

The interesting feature in this figure is the shape of the FOT line. Notice that at $T_{cp}$, the slope of the melting curve changes sign: above this temperature, the melting line has a negative slope $dB_m/dT$, which characterizes the melting curve at high temperatures where the FOT is governed by thermal fluctuations. Below this temperature, the melting line has a positive slope. This change in slope cannot be induced by thermal fluctuations because thermal fluctuations weaken as temperature decreases, and therefore the melting has to continue rising towards higher fields as the temperature decreases. Thus, the change in the curve direction must be induced by material disorder, i.e., *the extension of the FOT line beyond $T_{cp}$ is not just a matter of improved sensitivity but reveals a different FOT mechanism.*

These results strongly suggest that the FOT line does not terminate at $T_{cp}$, but continues as a disorder-driven FOT at lower temperatures. Moreover, this extended FOT line coincides with the second peak transition line $B_{sp}(T)$, as shown in Fig. 5b, which indicates that these two phenomena have a common origin – the reversible magnetization step is a thermodynamic signature of the transition, whereas the second magnetization peak is its dynamic characteristic reflecting the enhanced vortex pinning and slow dynamics in the disordered vortex phase. The revealed thermodynamic nature of the second peak transition is consistent with recent conclusions based on Josephson plasma resonance studies [21], transient measurements [22], magnetization studies [23] as well as several theoretical investigations [24]-[27]. These observations suggest that the melting transition and the second peak transition form a unified first-order transition that changes its character from a thermally induced transition to a disorder-driven one. Accordingly, the ordered Bragg glass phase [28],[29] is always destroyed through a FOT, either by thermal fluctuations at high temperatures, or by proliferation of disorder-induced dislocations at low temperatures, or by the combined effect of the two mechanisms.

An important open question is whether the two disordered states of the vortex matter, the liquid and the disordered phases, represent different thermodynamic phases or just a continuous slowdown of vortex dynamics. Recent numerical calculations lead to contradictory conclusions [24]-[27],[30]. We do not observe any sharp features along the depinning line $T_d$, which usually extends upward [31] from $T_{cp}$ and is believed to separate the two phases. A strongly first-order $T_d$ transition would require a sharp change in slope of $B_m(T)$ at $T_{cp}$, which we do not observe. Therefore, the $T_d$ line in BSCCO is unlikely to be a first-order transition; it could be either a continuous transition or a dynamic crossover. Similarly, in YBa$_2$Cu$_3$O$_7$ the structure of the phase diagram is controversial, and it is as yet unclear whether the second magnetization peak line merges with the melting line at the critical point [8] or at a lower field, and whether the line extending above the critical point is a second order transition [32].

While thermally induced melting of the vortex lattice is apparently present only in clean HTS, the disorder-driven FOT should be a more general phenomenon because it does not require enhanced thermal fluctuations. Indeed, the same point-disorder induced mechanism that leads to the second magnetization peak in HTS is apparently at the origin of the ubiquitous peak effect in the low-$T_c$ superconductors. Accordingly there is a growing recent evidence that the vortex lattice undergoes a disorder-driven FOT at the peak field of the peak effect in conventional superconductors like NbSe$_2$ [33]-[35] as well as in La$_{2-x}$Sr$_x$CuO$_4$ [36].

**5. Inverse Melting**

The shape of the FOT line, as shown in Fig. 5, reveals another important phenomenon. The region below $T_{cp}$, which is characterized by a positive slope, displays an unusual property. In contrast to the usual behavior where raising the temperature induces melting of a solid into a liquid, we observe an *inverse melting* process in which *heating* the vortex system induces transformation of a liquid or a *disordered* vortex phase into an *ordered* vortex lattice phase. This unexpected behavior has been recently found in polymeric systems [37],[38] as well as in some magnetic materials [39]. The thermodynamic nature of this inverse melting transition is apparently disorder-driven. This conclusion follows from trying to understand the shape of the FOT line in Fig. 5 in terms of competition between the different energies. At low temperatures, where thermal fluctuations have a minor role, the transition is disorder-driven and should be rather temperature independent [24]-[27],[40],[41]. Such a transition arises from the competition between the elastic energy of the lattice and the vortex pinning energy, where with increasing field the elastic energy of the lattice decreases relatively to the pinning energy, leading to a structural transition when the two energies become comparable. At intermediate temperatures, the transition *remains* disorder-driven, but thermal fluctuations reduce the pinning energy, resulting in an upturn in the transition line [40],[27].

It should be emphasized that inverse melting does not violate the thermodynamic requirements of larger entropy in the higher-temperature phase, but implies that the ordered vortex lattice has higher entropy than the low-temperature disordered phase. The entropy difference per unit volume between the disordered vortex phase and the vortex lattice can be calculated using the Clausius-Clapeyron relation

$$\Delta S = -\frac{\Delta B}{4\pi}\frac{dH_m}{dT}$$

The entropy difference per vortex per CuO layer is therefore

$$\Delta s = -\frac{\phi_0 d}{B_m}\frac{\Delta B}{4\pi}\frac{dH_m}{dT},$$

where $d$ is the interlayer spacing. The measured height of the discontinuous step $\Delta B$ and the associated $\Delta s$ are shown in Fig. 6 as a function of the melting temperature $T_m$. $\Delta B$ is rather constant in this range of temperatures. This observation resolves a previously reported inconsistency in the behavior near $T_{cp}$ – at a true critical point, $\Delta B$ is expected to decrease continuously to zero, whereas experimentally a rather constant $\Delta B$ was found to disappear abruptly at $T_{cp}$ [1]. This further suggests that $T_{cp}$ is not a critical point; therefore $\Delta B$ should not decrease to zero, and the previously reported disappearance of $\Delta B$ at $T_{cp}$ is an experimental limitation removed using the in-plane *ac* field. $\Delta s$, on the other hand, vanishes at $T_{cp}$ because $B_m(T)$ becomes horizontal when approaching $T_{cp}$. Below $T_{cp}$, $\Delta s$ becomes negative, meaning that the entropy is higher in the ordered lattice than in the disordered vortex phase. The surprising conclusion, considering the notion that a higher degree of order implies lower entropy, is that the lattice phase seems to be more disordered than the disordered vortex phase. Since the lattice is structurally more ordered as compared to the disordered phase, which has a high concentration of dislocations, the extra entropy must arise from additional degrees of freedom. Therefore, even though there are no dislocations and the time averaged vortex positions retain a rather long-range order, the thermal fluctuations on short scales are apparently larger in the ordered lattice than in the entangled disordered phase.

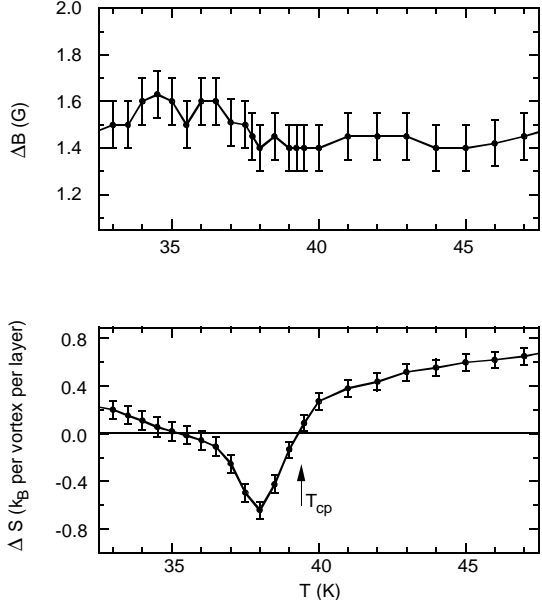

Fig. 6 The measured height of the discontinuous step *DB* and the entropy difference *Ds* per vortex per CuO layer

It should be noted that *DB* is relatively high compared to the step height that was measured at higher temperatures in this crystal. $H_{ac\perp}$ was found to increase the height of the discontinuity, and this effect becomes more significant when increasing the amplitude of $H_{ac\perp}$. This raises the question of whether $H_{ac\perp}$ reveals the true height of *DB*, which was suppressed due to pinning, or whether it slightly distorts its value. Further verification by additional methods is required in order to clarify this point.

## 6. Conclusions

We have found that the FOT extends beyond the critical point, which implies that the previously reported termination of the FOT at $T_{cp}$ does not reflect a real critical point, but rather an experimental limitation. Detailed mapping of the transition lines on the *B-T* phase diagram has shown that this FOT changes its direction at $T_{cp}$ and coincides with the second peak transition at low temperatures. These observations imply that the second magnetization peak is a first-order transition. Accordingly, the quasi-ordered vortex lattice phase is destroyed through a unified first-order transition that changes its character from a thermal melting transition to a disorder induced one. At intermediate temperatures, this transition is found to display unusual inverse melting behavior, where a disordered vortex phase transforms into an ordered lattice with increasing temperature. Paradoxically, the structurally ordered lattice has larger entropy than the disordered phase.

## Acknowledgments

This work was supported by the Israel Science Foundation - Center of Excellence Program, by the Ministry of Science, Israel, by Minerva Foundation, Germany, and by the Grant-in-Aid for Scientific Research from the Ministry of Education, Science, Sports and Culture, Japan. DEF acknowledges the support by the Koshland Fellowship and RFBR grant. PK and ML acknowledge support of the Dutch Foundation FOM. EZ acknowledges support of the Mitchell Research Fund.